\documentclass[a4paper]{article}

\usepackage{icrc2013}
\usepackage{amsmath}
\usepackage{amssymb}
\usepackage{enumitem}
\usepackage{wasysym}
\usepackage{upgreek}

\title{Status of the Silicon Photomultiplier Telescope FAMOUS for the Fluorescence Detection of UHECRs}

\shorttitle{Status of the SiPM Telescope FAMOUS for the Detection of UHECRs}

\authors{
  Tim Niggemann$^{1}$,
  Pedro Assis$^{2}$,
  Pedro Brogueira$^{2}$,
  Antonio Bueno$^{3}$,
  Hans Michael Eichler$^{1}$,
  Miguel Ferreira$^{2}$,
  Thomas Hebbeker$^{1}$,
  Markus Lauscher$^{1}$,
  Lu\'{i}s Mendes$^{2}$,
  Lukas Middendorf$^{1}$,
  Sergio Navas$^{3}$,
  Christine Peters$^{1}$,
  M\'{a}rio Pimenta$^{2}$,
  Angel Ruiz$^{3}$,
  Johannes Schumacher$^{1}$,
  Maurice Stephan$^{1}$
}

\afiliations{
  $^1$ III. Phys. Inst. A, RWTH Aachen University, Aachen, Germany \\
  $^2$ Laborat\'{o}rio de Instrumenta\c{c}\~{a}o e F\'{i}sica Experimental de Part\'{i}culas, Lisbon, Portugal \\
  $^3$ Departamento de Fisica Te\'{o}rica y del Cosmos, Universidad de Granada, Spain\\ ICRC 2013, Rio de Janeiro, Brazil \\
}

\email{niggemann@physik.rwth-aachen.de\\http://www.physik.rwth-aachen.de/institute/institut-iiia/forschung/auger/famous}

\abstract{An established technique for the measurement of ultra-high-energy-cosmic-rays is the detection of the fluorescence light induced in the atmosphere of the Earth, by means of telescopes equipped with photomultiplier tubes.
Silicon photomultipliers (SiPMs) promise an increase in the photon detection efficiency which outperforms conventional photomultiplier tubes. In combination with their compact package, a moderate bias voltage of several ten volt and single photon resolution, the use of SiPMs can improve the energy and spatial resolution of air fluorescence measurements, and lead to a gain in information on the primary particle. Though, drawbacks like a high dark-noise-rate and a strong temperature dependency have to be managed.
FAMOUS is a refracting telescope prototype instrumented with 64 SiPMs of which the main optical element is a Fresnel lens of $549.7\,\text{mm}$ diameter and $502.1\,\text{mm}$ focal length. The sensitive area of the SiPMs is increased by a special light collection system consisting of Winston cones. The total field of view of the telescope is approximately $12\,^\circ$. The frontend electronics automatically compensates for the temperature dependency of the SiPMs and will provide trigger information for the readout. Already for this prototype, the Geant4 detector simulation indicates full detection efficiency of extensive air showers of $E=10^{18}\,\text{eV}$ up to a distance of $6\,\text{km}$. 
We present the first working version of FAMOUS with a focal plane prototype providing seven active pixels.}


\begin{document}
\maketitle

\section{Fluorescence Detection of High Energy Cosmic Rays}
The fluorescence detection technique is employed in many experiments for the measurement of high energy cosmic rays with energies $E>10^{17}\,\text{eV}$ as e.g{.} in the Pierre Auger Observeratory located in Argentina \cite{bib:abraham2010}.

As a (primary) particle penetrates the atmosphere of the Earth, a shower of secondary particles is produced continuously until the number of particles reaches a maximum at a slant depth $X_\text{max}$ and the shower dies out. The number of particles along the axis of the extensive air shower (EAS) can be described by the Gaisser-Hillas-parametrisation \cite{bib:bluemer2009}.

The secondary particles produced in the EAS excite the nitrogen molecules of the air. When the molecules transit to the ground state, fluorescence light in the regime of $290\,\text{nm}$ to $440\,\text{nm}$\footnote{The most dominant transition is at $\lambda=337\,\text{nm}$ \cite{bib:arqueros2008}.} is emitted isotropically whereby the number of photons is proportional to the energy deposit $E_{\text{dep}}^{\text{tot}}$\cite{bib:arqueros2008}{:}
\begin{equation}
	\frac{\text{d}^{2}N_{\gamma}}{\text{d}X\text{d}\lambda}=Y(\lambda,P,T,u)\cdot\frac{\text{d}E_{\text{dep}}^{\text{tot}}}{\text{d}X}\qquad\text{.}\label{eq:fluorescence-yield}
\end{equation}
The light yield $Y$ depends on the wavelength $\lambda$, the atmospheric pressure $P$ , the temperature $T$ and the humidity $u$. In dry air with $P=1000\,\text{hPa}$ and $T=300\,\text{K}$, the absolute value is $Y(\lambda=337\,\text{nm}) \approx 5\,\text{MeV}^{-1}$ and $\int Y \text {d}\lambda \approx 19\,\text{MeV}^{-1}$ \cite{bib:arqueros2008}. E.g. for a $6\,\text{km}$ distant air shower with an energy of $E=10^{18}\,\text{eV}$, approx{.} $10^8$ photons arrive at a telescope with an aperture of $500\,\text{mm}$ diameter and a field of view of $12\,^\circ$. Using the time of arrival information of the light, it is possible to obtain the Gaisser-Hillas-shape and $X_\text{max}$ of the EAS which gives access to the energy, direction and mass of the primary particle.

The telescope FAMOUS (``\underline{F}irst \underline{A}uger \underline{M}ulti pixel photon counter camera for the \underline{O}bservation of \underline{U}ltra-high-energy-cosmic-ray air \underline{S}howers'') is a compact refractor with a camera built-up of novel photodetectors: Silicon photomultipliers (SiPMs). The aim of this effort is to demonstrate the feasibility of SiPMs for the fluorescence detection of cosmic rays which would enable less expensive and possibly more accurate measurements of cosmic rays. 

\section{Silicon Photomultipliers}

\begin{figure}[t]
  \centering
  \includegraphics[height=0.481\textwidth]{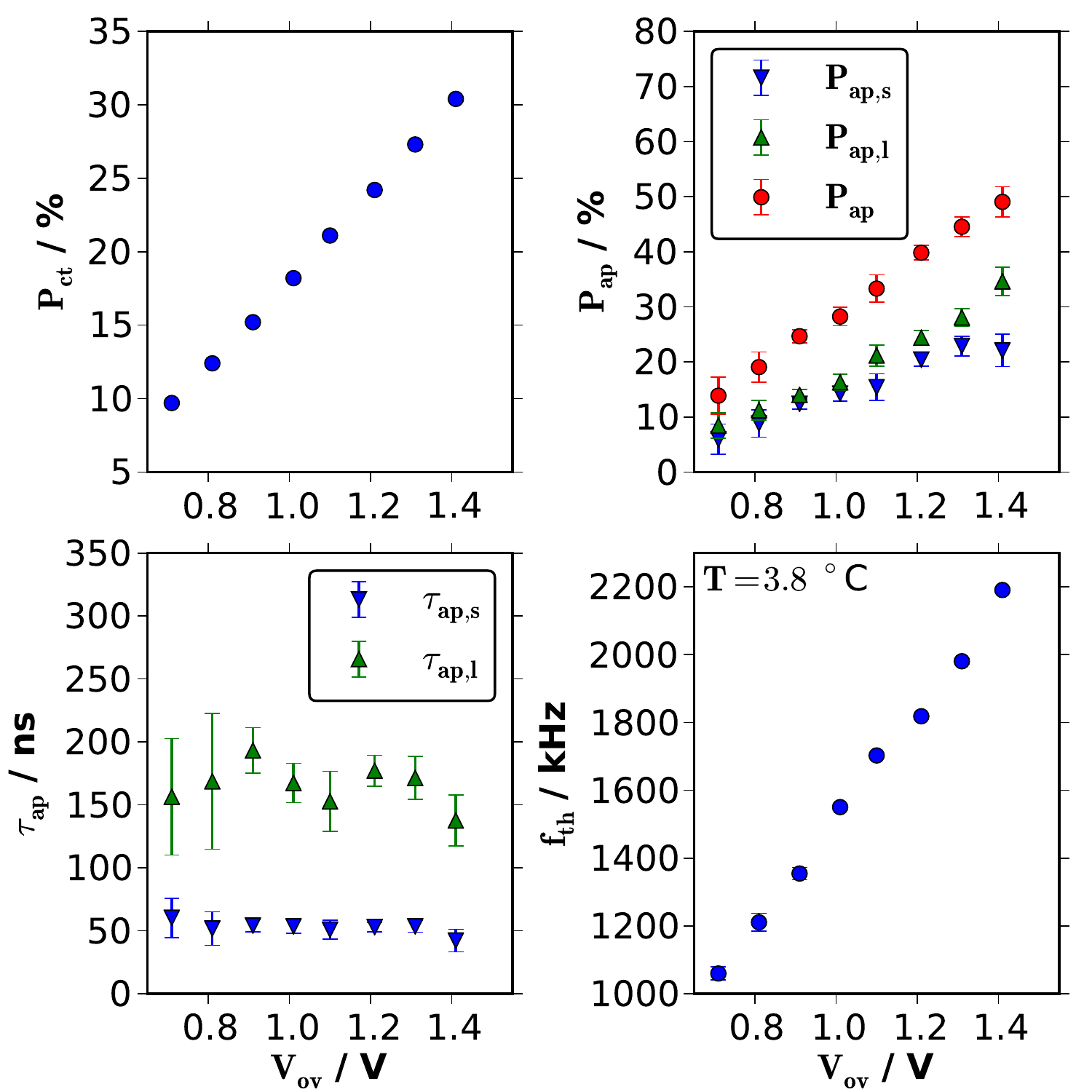}
  \caption{Measurement of the SiPM operation parameters crosstalk- ($P_\text{ct}$) and afterpulse-probability ($P_\text{ap,s}$, $P_\text{ap,l}$, combined probability $P_\text{ap}$), the afterpulsing time constants ($\tau_\text{ap,s}$ and $\tau_\text{ap,l}$) and the thermal noise rate ($f_\text{th}$) as a function of the over-voltage $V_\text{ov}$ for a Hamamatsu S10362-33-100C SiPM.
The ambient temperature is $T = \left(3.76 \pm 0.03\right)\,^\circ\text{C}$.}
  \label{fig:overvoltage}
\end{figure}

Silicon photomultipliers (SiPMs) are semiconductor photodetectors which are built-up of hundreds to thousands of small avalanche photodiodes, referred to as cells, operated in Geiger-mode. Thus, each cell is capable of single photon detection and the dynamic range of the whole device is defined by the number of cells situated on the SiPM ($N_\text{cells}$). Currently, cell pitches between typ{.} $25\,\upmu\text{m}$ to $100\,\upmu\text{m}$ are commercially available (see e.g{.} \cite{bib:hamamatsu2009}). For very short photon pulses, the number of triggered cells $N_\text{trig}$ can be expressed as a function of the number of incoming photons $N_\lambda$
\begin{equation}
 N_\text{trig}\left( N_\lambda \right)=N_\text{cells}\left(1 - e^{-N_\lambda\,PDE / N_\text{cells}} \right)
\end{equation}
with the photon detection efficiency $PDE$ which itself depends on the wavelength of the photons. 

Due to the Geiger-mode, the gain of SiPMs is typ{.} in the order of $10^6$ \cite{bib:renker2009}. The gain depends linearly on the voltage excess $V_\text{ov}$ of the applied bias voltage $V_\text{bias}$ above the breakdown voltage $V_\text{break}$ (typ{.} $70\,\text{V}$)
\begin{equation}
 V_\text{ov} = V_\text{bias} - V_\text{break} \qquad \text{.}
\end{equation}
By increasing the over-voltage, the $PDE$ can also be increased but at the expense of increasing noise. 

The SiPMs used in FAMOUS are of type Hamamatsu-S10985-100C which in turn is composed of an array of four Hamamatsu S10362-33-100C SiPMs. The photon detection efficiency at $\lambda=420\,\text{nm}$ and $V_\text{ov}=1.3\,\text{V}$ is $PDE=\left(33.1 \pm 0.8\pm 5\right)\,\text{\%}$ \cite{bib:lauscher2012}.

\subsection{Noise Phenomena}
Even if operated in the dark, cell breakdowns can happen accidentally by thermal excitation. A typical thermal noise rate is
\begin{equation}
 f_\text{th} = 1\,\text{kHz}\cdot N_\text{cells}\qquad
\end{equation}
at room temperature. Additionally, SiPMs are subject to two kinds of correlated noise: optical crosstalk and afterpulsing.

\paragraph{Optical Crosstalk}
Charge carriers of the avalanche during the cell breakdown can recombine (probability approx{.} $10^{-5}$ per avalanche electron) to a photon which itself can cause an avalanche breakdown in a neighbouring cell. The probability for this process is $P_\text{ct}\approx10\ldots30\,\text{\%}$ depending on the over-voltage.

\paragraph{Afterpulsing}
Also during the avalanche, charge carriers may be trapped at impurities of the silicon substrate of the SiPM and be released after a characteristic time $\tau$ which usually exceeds the time needed to replenish the avalanche zone of the SiPM cell. The released charge carriers can cause the cell to break down again. Measurements of Hamamatsu devices revealed a short (with $\tau_\text{ap,s}\approx50\,\text{ns}$) and a long time constant (with  $\tau_\text{ap,l}\approx170\,\text{ns}$) with a combined afterpulsing probability (combined probabilities $P_\text{ap,s}$ and $P_\text{ap,l}$ for the short and long time constant) of $P_\text{ap}\approx10\ldots40\,\text{\%}$.


The SiPM properties discussed above are presented in figure \ref{fig:overvoltage} for a Hamamatsu S10362-33-100C SiPM as a function of $V_\text{ov}$.

A dramatic suppression of thermal and correlated noise is expected with the release of future devices \cite{bib:hamamatsu2013}.

\section{Baseline Design of FAMOUS}

\begin{figure}[t]
  \centering
  \includegraphics[height=0.481\textwidth]{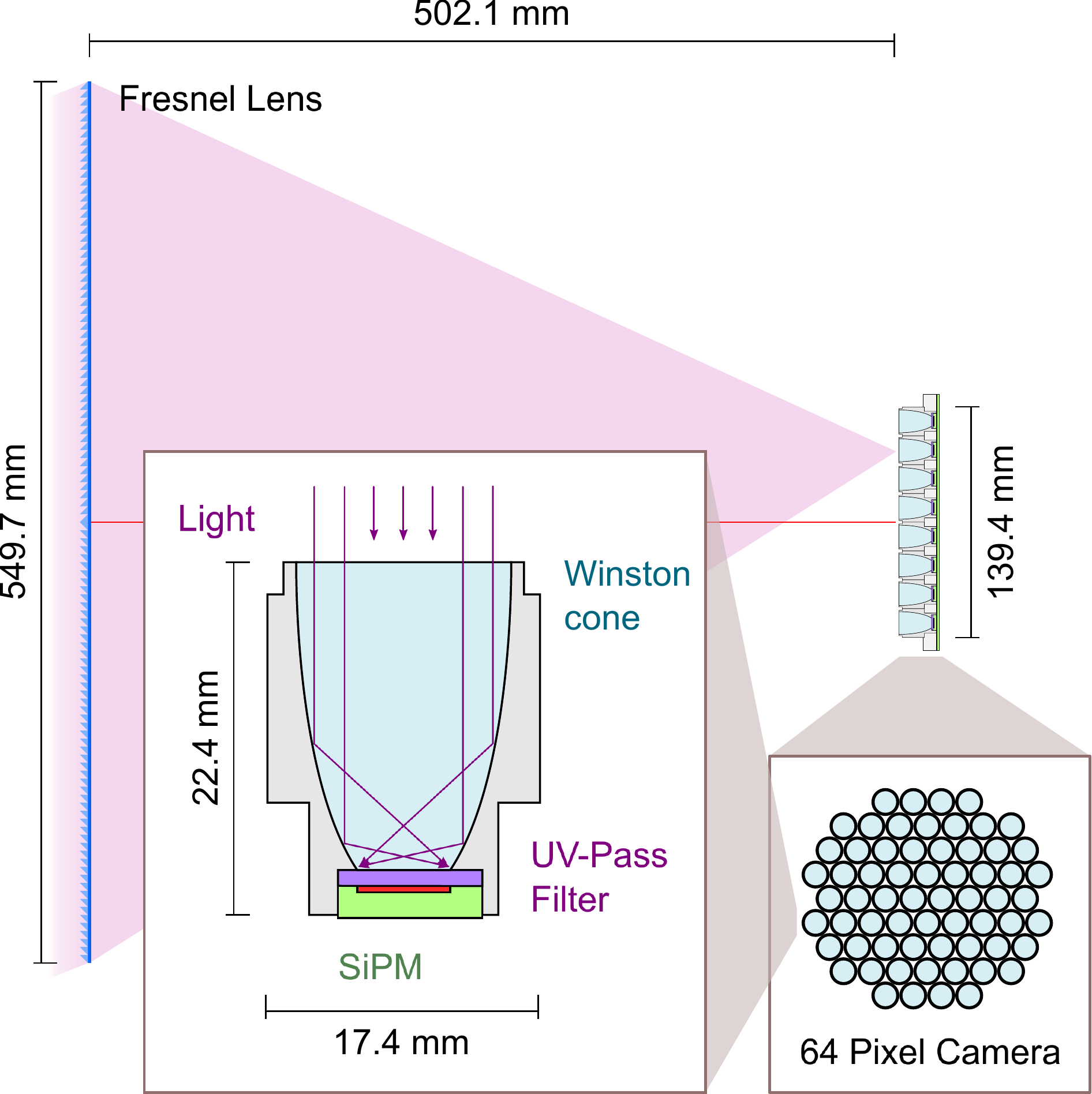}
  \caption{Baseline design of FAMOUS. The incoming light is concentrated by the $549.7\,\text{mm}$ diameter Fresnel lens with $f=502.1\,\text{mm}$ focal length onto a camera with $64$ SiPMs which are each located behind a UV-pass filter and a light concentrator, the Winston cone. The pixels are packed in a hexagonal layout. The illustration is true to scale.}
  \label{fig:baseline}
\end{figure}

\begin{figure}[t]
  \centering
  \vspace{0.75em}
  \includegraphics[width=0.481\textwidth]{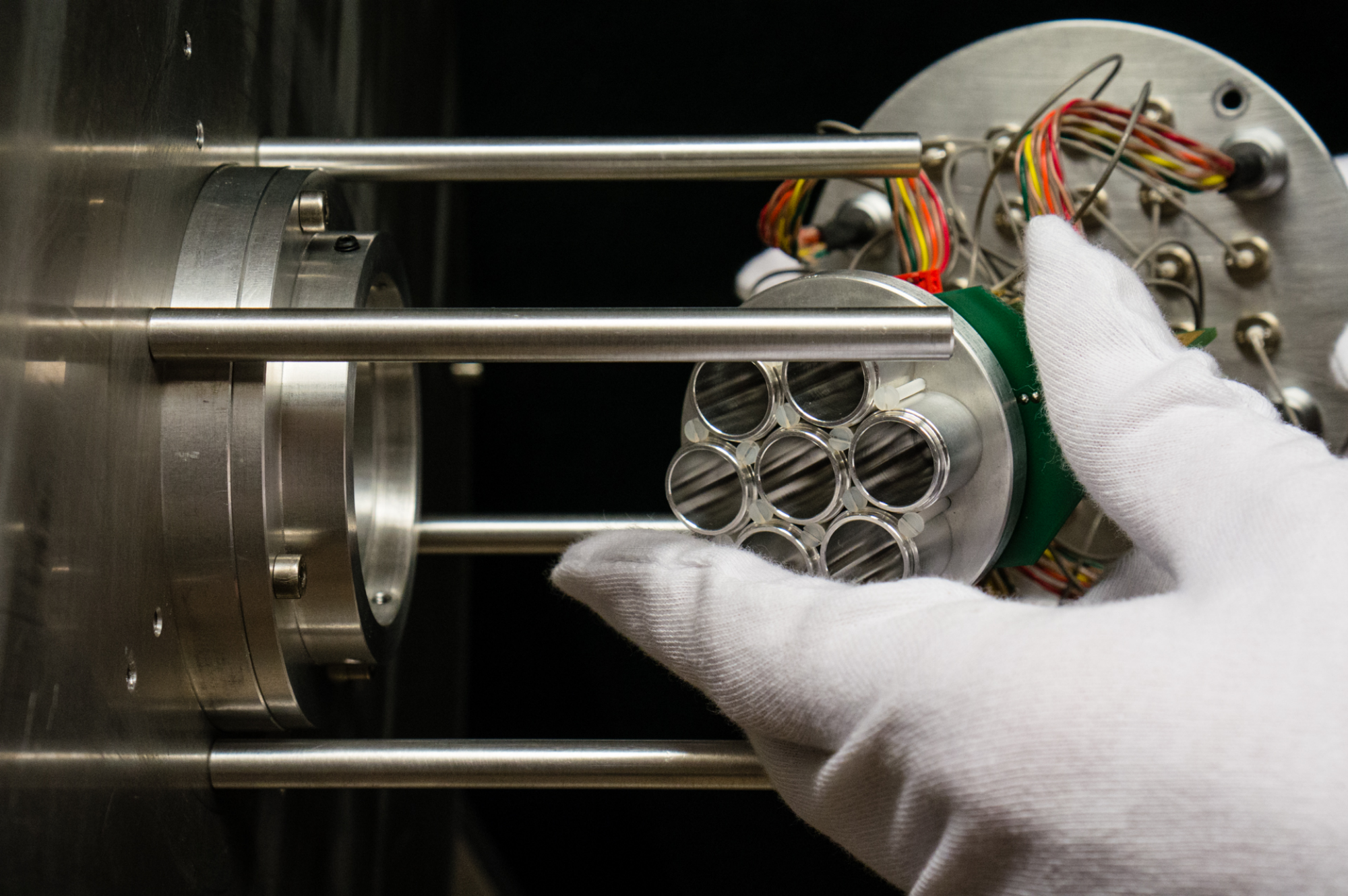}
  \caption{Photo of the assembly of the FAMOUS$^\text{SEVEN}$ prototype focal plane with seven active pixels. The SiPMs (Hamamatsu-S10985-100C) are located behind the circular Winston cones.}
  \label{fig:assembly}
\end{figure}

\begin{figure}[t]
  \centering
  \vspace{0.75em}
  \includegraphics[width=0.481\textwidth]{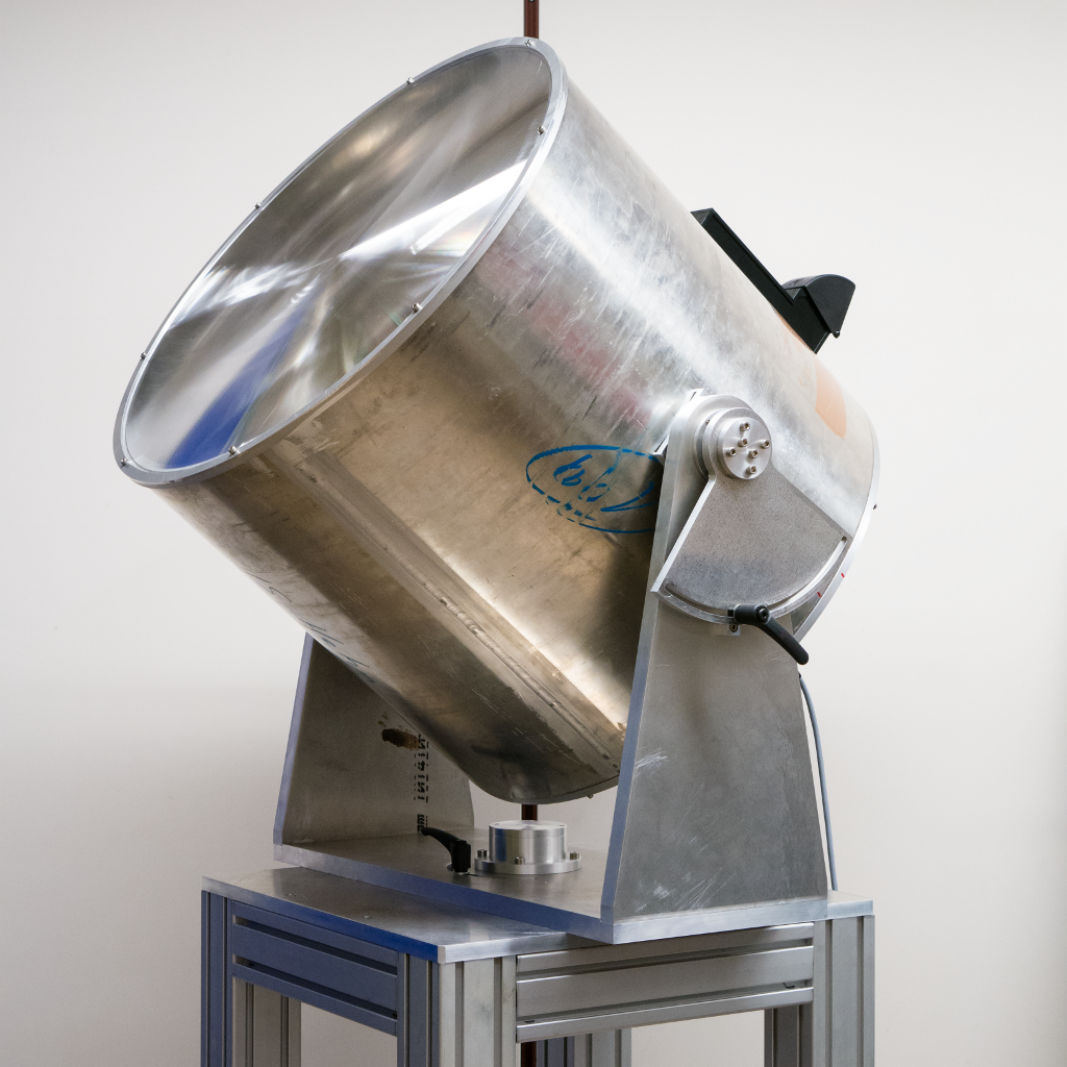}
  \caption{Photo of the fully assembled FAMOUS$^\text{SEVEN}$ prototype with the Fresnel lens attached to the telescope tube which is fixed onto an altitude-azimuth mount.}
  \label{fig:assembly2}
\end{figure}

FAMOUS is a refracting telescope. A commercial Fresnel lens \cite{bib:fresnel2013} with an aperture of $d=549.7\,\text{mm}$ diameter (design goal $d=510\,\text{mm}$) and a focal length of $f=502.1\,\text{mm}$ (design goal $f=510\,\text{mm}$) is used to concentrate the light onto the camera composed of 64 circular pixels arranged in a hexagonal layout. Each pixel has a circular field of view of $1.5\,^\circ$. A schematic of the baseline design is presented in figure \ref{fig:baseline}.

\subsection{Fresnel Lens}

To minimise weight and light loss due to absorption, a Fresnel lens is divided into concentric, annular sections while the bulk material of the sections is reduced to the minimum possible. The Fresnel lens of FAMOUS has $10$ sections per millimeter and is made of UV transparent acrylic \cite{bib:fresnel2011,bib:niggemann2012}. The overall transmission has been simulated to $T\approx70\,\text{\%}$ over the whole fluorescence light regime. 

The image quality has also been determined by ray-tracing simulations to $R_{90}=1.88\,\text{mm}$ \cite{bib:niggemann2012}. $R_{90}$ signifies $90\,\text{\%}$ of encircled energy in the image plane for incidence of parallel light, here perpendicularly to the lens. Recent measurements resulted in $R_{90}=2.89\,\text{mm}$ which is good enough for the application in FAMOUS and reasonably matches the simulations considering that the simulations did not account for surface imperfections and shape deviations, and the incoming light used for the measurement was not perfectly parallel.

\subsection{Camera Pixel}

Each pixel of FAMOUS consists of a light concentrator made of polished aluminium (Winston cone \cite{bib:winston2005}), a $1\,\text{mm}$ thin UV-pass filter (Schott UG-11 \cite{bib:schott2008}) and a Hamamatsu-S10985-100C SiPM, i.e. an array of four $3 \times 3\,\text{mm}^2$ Hamamatsu S10362-33-100C SiPMs yielding a total area of $6\times6\,\text{mm}^2$, and a cell pitch of $100\,\upmu\text{m}$.

The Winston cone is a non-imaging parabolic concentrator with a circular entrance of $r_1=6.71\,\text{mm}$ radius and an exit radius of $r_2=3\,\text{mm}$. Thus, up to $\theta_\text{max}=26.6\,^\circ$ incidence angle, over $90\,\text{\%}$ of the incoming light is transmitted to the exit \cite{bib:niggemann2012}. Due to manufacturing processes, the Winston cones of FAMOUS are circular, but in principle rectangular or even hexagonal shapes are also possible.

\subsection[FAMOUS-7]{FAMOUS$^\text{SEVEN}$}

The first version of the focal plane features seven pixels to test the feasibility of the mechanical design (c.f. figures \ref{fig:assembly} and \ref{fig:assembly2}) and the scalability of the amplifier electronics developed in-house. 

A dedicated readout with $64$ individual channels and a programmable FPGA chip for triggering developed in Lisbon is currently being tested and will be ready soon for the implementation in the $64$ pixel version of FAMOUS.

\section{Expected Performance}

\begin{figure}[t]
  \centering
  \includegraphics[width=0.481\textwidth]{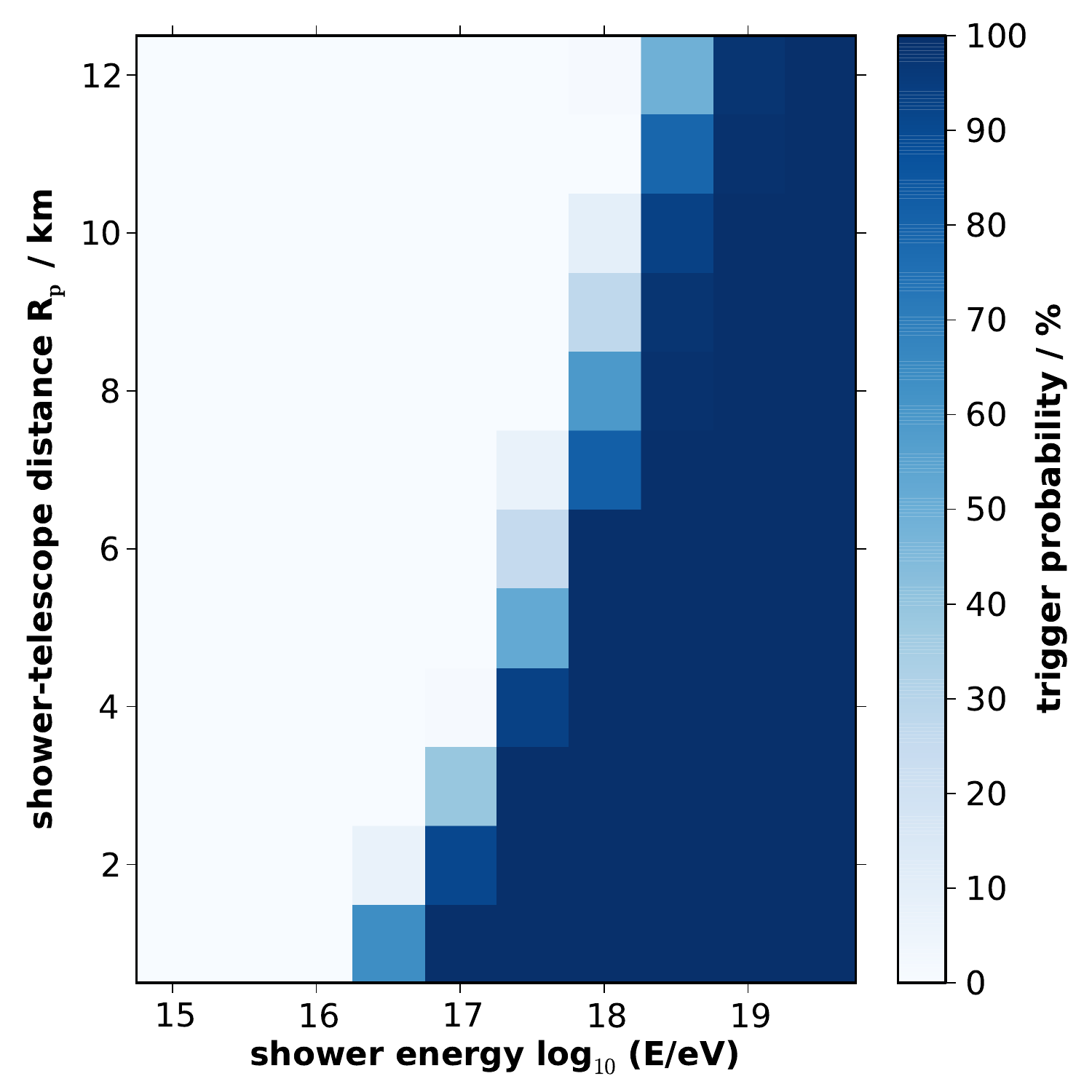}
  \caption{Simulated trigger probability for vertical EAS as a function of the shower-to-telescope distance and the shower energy.}
  \label{fig:triggerprobability}
\end{figure}

To evaluate the baseline design of FAMOUS, a dedicated Geant4 \cite{bib:agostinelli2003} ray-tracing-simulation has been developed \cite{bib:niggemann2012,bib:stephan2011,bib:lauscher2012famous}. The air shower simulations have been carried out with CONEX \cite{bib:bergmann2007}.

The trigger probability determined by the detector simulation is presented in figure \ref{fig:triggerprobability}. It can be used as a benchmark quantity for the evaluation of the performance of the real telescope. An EAS has been considered as successfully triggered if the fluorescence signal exceeded the noise in a pixel of FAMOUS by a certain threshold. The noise is dominated by the night-sky-brightness
. The night-sky-background-radiance was measured in Aachen, Germany, to $L\lesssim1.9\cdot10^{12}\,\text{m}^{-2}\text{s}^{-1}\text{sr}^{-1}$ in the fluorescence light regime \cite{bib:stephan2011}. Thus, a $10^{18}\,\text{eV}$ shower is detectable at a shower-to-telescope distance of $6\,\text{km}$ (at full efficiency) and at a rate of a few ten EAS in total per measurement night. Since SiPMs are invulnerable to too much light and the image quality of the Fresnel lens is sufficient, measurements at full moon or twilight may become possible.

\section{First Test Measurements}

\begin{figure}[t]
  \centering
  \includegraphics[width=0.481\textwidth]{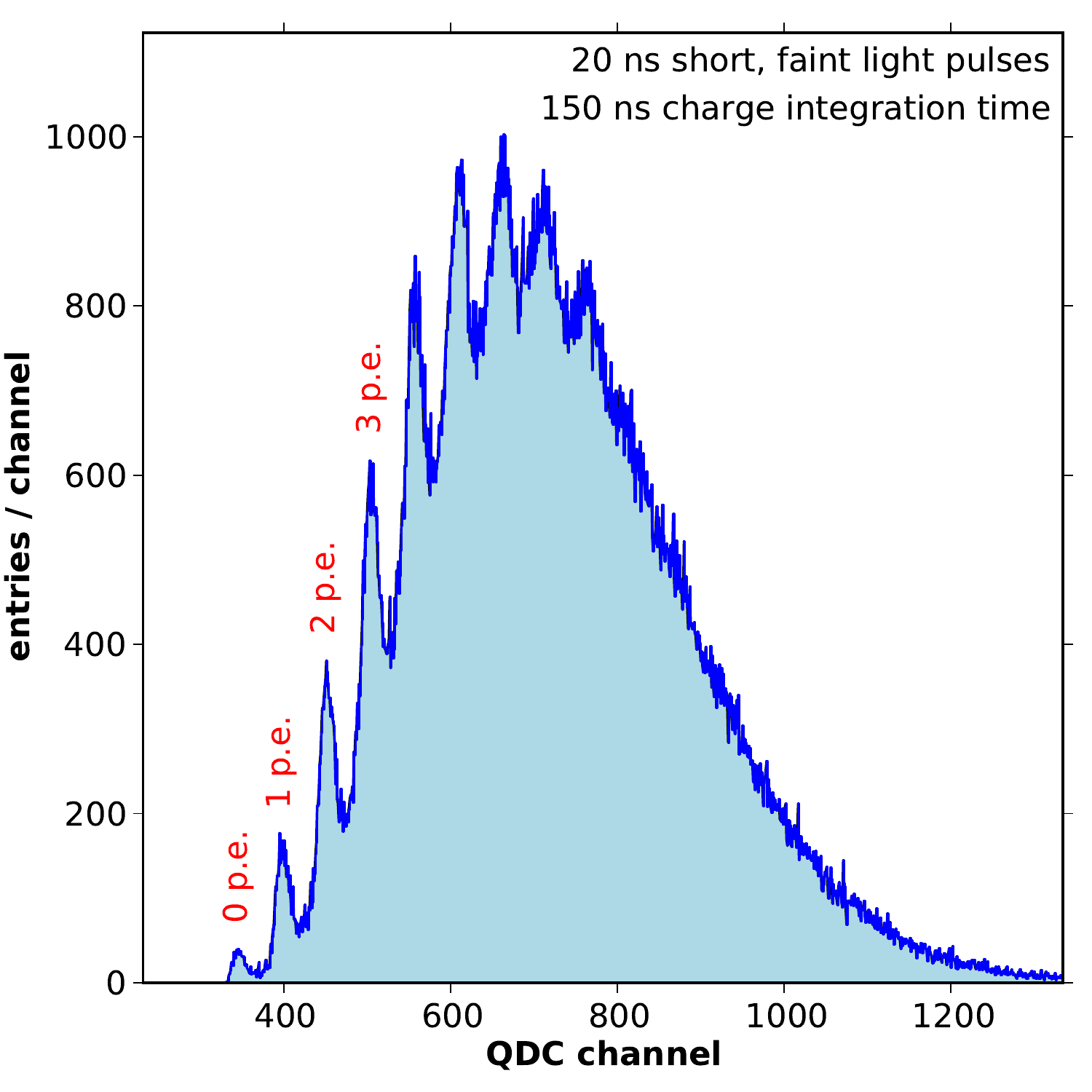}
  \caption{Charge spectrum of a single pixel of FAMOUS when illuminated by pulsed LED light (peak wavelength $400\,\text{nm}$) measured with a QDC. The single peaks of the photon equivalents can be easily distinguished.}
  \label{fig:fingers}
\end{figure}

Upon completion of the assembly of FAMOUS$^\text{SEVEN}$, first test measurements were performed. In figure \ref{fig:fingers}, a charge spectrum of a single pixel signal read out by the amplifier electronics of FAMOUS$^\text{SEVEN}$ is presented. The data have been taken with a pulsed LED whose peak wavelength is at $400\,\text{nm}$. The pulse width was $20\,\text{ns}$ and the integration time of the charge-to-digital-converter (QDC) $150\,\text{ns}$. The photon equivalent peaks are clearly distinguishable which enables characterisation studies and is necessary for the over-voltage respectively gain calibration of the pixels.

\section{Summary \& Outlook}

The refractive optics of the air fluorescence telescope prototype FAMOUS consists of a commercial Fresnel lens with $d=549.7\,\text{mm}$ nearly equal to the focal length $f=502.1\,\text{mm}$. To match an effective field of view of $1.5\,^\circ$ per pixel, the sensitive area of the Hamamatsu S10985-100C SiPM (i.e{.} an array of four $3\times3\,\text{mm}^2$ SiPMs) is increased by a special light concentrator, a Winston cone. The Winston cone has an entrance radius of $r_{1}=6.7\,\text{mm}$. The focal plane of FAMOUS features $64$ hexagonally collocated pixels in total.

CONEX air shower and detailed Geant4 detector simulations have been performed, and successfully demonstrated the \mbox{ability} of FAMOUS to detect air showers even with its \mbox{small} aperture considering contemporary air fluorescence telescopes. 

The first version with seven pixels, FAMOUS$^\text{SEVEN}$, is fully assembled and operational. Later this year, the focal plane of FAMOUS will be extended to $64$ pixels as soon as the dedicated electronics are finished and programmed.

\vspace{1em}
\footnotesize{{\bf Acknowledgment: }{This work is funded by the 
European astroparticle physics network ASPERA, the German Federal Ministry of Education and Research BMBF, the Programa Operacional Factores de Competitividade Compete, the Funda\c{c}\~{a}o para a Ci\^{e}ncia e a Tecnologia, the Fundo Europeu de Desenvolvimento Regional and the Quadro de Referência Estrat\'{e}gica Nacional. Additionally, the authors would like to thank the mechanical and electronics workshops in Aachen, Granada and Lisbon and the CMS working group Aachen.}}

\end{document}